\documentclass[runningheads]{llncs}
\usepackage[T1]{fontenc}
\usepackage{enumitem}
\usepackage{graphicx}
\usepackage{times}
\usepackage{latexsym}
\usepackage{booktabs}
\usepackage{float}
\usepackage{enumerate}
\usepackage[utf8]{inputenc}
\usepackage{microtype}
\usepackage{inconsolata}
\usepackage{xcolor}

\begin{document}
\title{FinTeam: A Multi-Agent Collaborative Intelligence System for Comprehensive Financial Scenarios}

\titlerunning{FinTeam: Multi-Agent Financial Intelligence System}
%
\author{
Yingqian Wu\inst{1} \and
Qiushi Wang\inst{1} \and
Zefei Long\inst{1} \and
Rong Ye\inst{1} \and
Zhongtian Lu\inst{1} \and
Xianyin Zhang\inst{1} \and
Bingxuan Li\inst{1} \and
Wei Chen\inst{2} \and
Liwen Zhang\inst{3} \and
Zhongyu Wei\inst{1,4,5}\thanks{Corresponding author.}
}
\authorrunning{Y. Wu et al.} 

\institute{%
  School of Data Science, Fudan University, China \and
  School of Software Engineering, Huazhong University of Science and Technology, China \and
  Shanghai University of Finance and Economics, China \and
  Shanghai Innovation Institute, China \and
  Research Institute of Intelligent Complex Systems, Fudan University, China\\[4pt]
  \email{%
    \{wuyq23,qswang23,zflong23,yer23,ztlu22,xianyinzhang22,bxli16\}@m.fudan.edu.cn,\\
    lemuria\_chen@hust.edu.cn, zhang.liwen@shufe.edu.cn, zywei@fudan.edu.cn%
  }%
}

%
%
\maketitle              
\begin{abstract}
Financial report generation tasks range from macro- to micro-economics analysis, also requiring extensive data analysis. Existing LLM models are usually fine-tuned on simple QA tasks and cannot comprehensively analyze real financial scenarios. 
Given the complexity, financial companies often distribute tasks among departments. 
Inspired by this, we propose FinTeam, a financial multi-agent collaborative system, with a workflow with four LLM agents: \textit{document analyzer}, \textit{analyst}, \textit{accountant}, and \textit{consultant}.
We train these agents with specific financial expertise using constructed datasets.
We evaluate FinTeam on comprehensive financial tasks constructed from real online investment forums, including macroeconomic, industry, and company analysis. The human evaluation shows that by combining agents, the financial reports generate from FinTeam achieved a 62.00\% acceptance rate, outperforming baseline models like GPT-4o and Xuanyuan. 
Additionally, FinTeam's agents demonstrate a 7.43\% average improvement on FinCUGE and a 2.06\% accuracy boost on FinEval. 
Project is available at https://github.com/FudanDISC/DISC-FinLLM/.

\end{abstract}
\section{Introduction}
\begin{figure}[t]
    \centering
    \includegraphics[width=0.6\columnwidth]{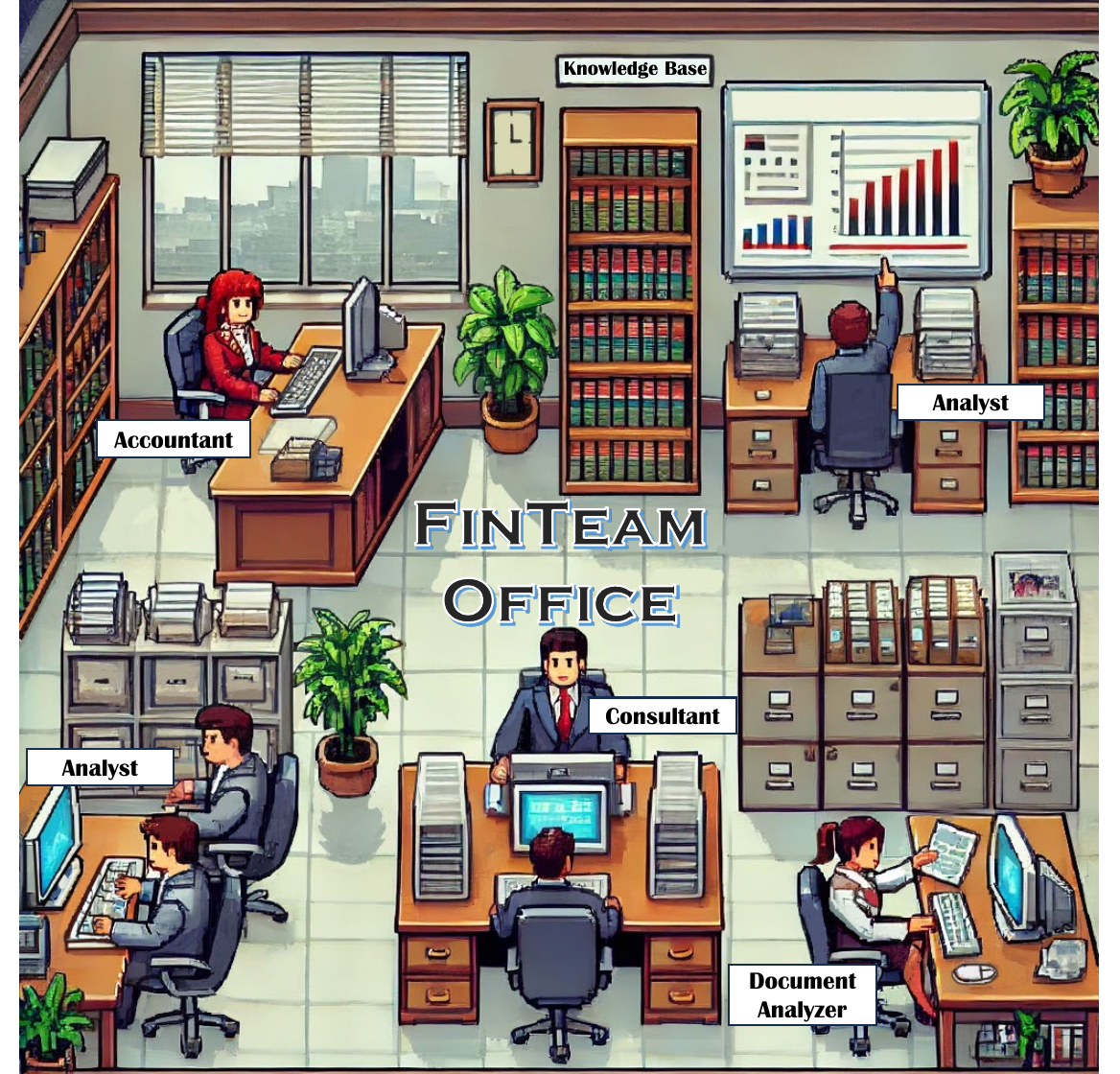}
    \caption{Inspired by the financial companies that assign tasks to specialized teams, FinTeam distributes financial tasks among the \textit{document analyzer}, \textit{analyst}, \textit{accountant}, and \textit{consultant} agents, enabling a more efficient and sophisticated process.}
    \label{fig:finsys}
\end{figure}

The financial industry presents unique challenges and opportunities for Large Language Models (LLMs).
Recently, domain-specialized LLMs have achieved notable progress in several vertical fields,
such as legal reasoning~\cite{yue2024lawllm} and medical consultation~\cite{bao2023discmedllm}.
While LLMs like BloombergGPT~\cite{wu2023bloomberggpt} and XuanYuan~\cite{zhang2023xuanyuan} perform well on routine tasks, they struggle with complex, multifaceted financial problems. Single LLM calls often lack the precision needed for detailed financial analysis. In practice, as shown in Figure~\ref{fig:finsys}, financial tasks are typically handled by specialists, highlighting the potential of decomposing tasks into sub-tasks managed by dedicated agents.

In response to these challenges, we introduce FinTeam, a financial intelligence system composed of multiple collaborating LLM agents, each designed to address specific scenarios in finance. We focus on three key financial scenarios: macroeconomic analysis, industry analysis, and company analysis. These scenarios encompass a range of tasks, from assessing broad economic trends to detailed evaluations of individual companies. Within this framework, four specialized LLM agents collaborate to handle their respective tasks and provide comprehensive solutions.

The four LLM agents focus on processing financial texts such as news and company reports, performing real-time material analysis based on knowledge bases, using computational tools to achieve financial numerical calculation, and professionally answering a variety of financial questions.

To demonstrate FinTeam's effectiveness, we conduct extensive evaluations on both the collaborative system and individual agents. Using a dataset of 150 real investor queries spanning the three scenarios, we compare FinTeam's performance to that of baseline models, including Qwen2.5-7B-Instruct, GPT-4o, ChatGLM3-6B and Xuanyuan-13B. The results, evaluated by GPT-4o, show that FinTeam has an overall score of 4.86 (out of 5), which is significantly better than the other models by 0.03 to 0.82 points. Furthermore, human preference evaluations confirm FinTeam's real-world relevance, with a winning rate of 62.00\%. At the same time, the performance of each agent is also significantly better than the other models in each evaluation.

In summary, our contributions are as follows:
\begin{itemize}[itemsep=1pt, leftmargin=10pt, parsep=0pt, topsep=1pt]
    \item We design FinTeam, a financial intelligence system comprising multiple collaborating LLM agents: the \textit{analyst}, \textit{document analyzer}, \textit{accountant}, and \textit{consultant}. Each agent addresses specific challenges in various financial scenarios. 
    \item To strengthen their capabilities, we develop the agent training dataset used for fine-tuning. Extensive evaluations demonstrate FinTeam's effectiveness and the value of the dataset. 
    \item Unlike prior evaluations that focus on simple financial tasks, we concentrate on three comprehensive real-world application scenarios: macroeconomic analysis, industry analysis, and company analysis. Both automated evaluations and human assessments demonstrate the effectiveness of FinTeam's workflow.
\end{itemize}

\begin{figure}[t]
    \centering
    \includegraphics[width=0.95\columnwidth]{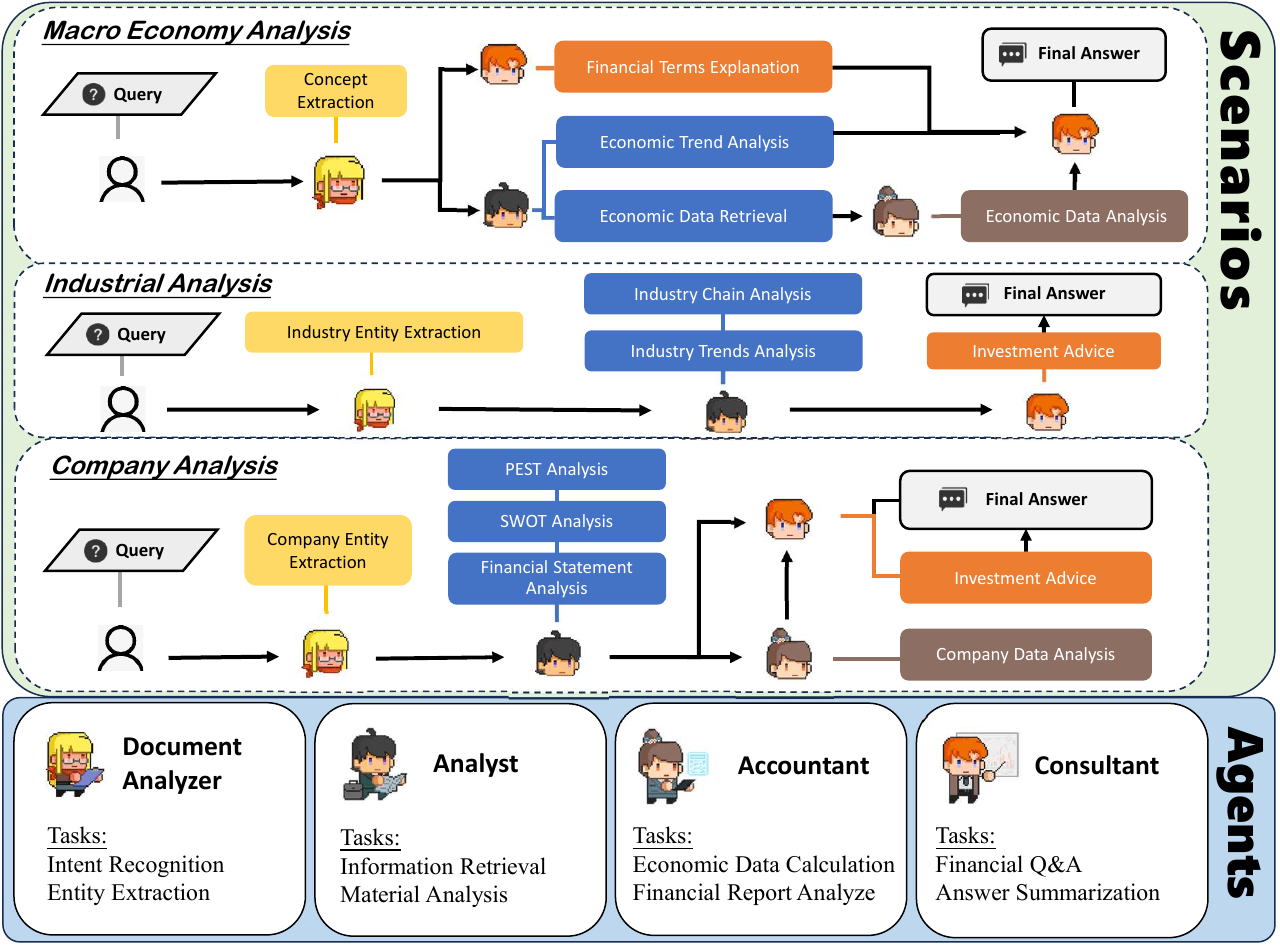}
    \caption{Overview of FinTeam Multi-Agent Collaborative Intelligence System.}
    \label{fig:finllm}
\end{figure}

\section{Related Work}
\subsection{LLMs for Finance}
\label{sec:2.1}

Large Language Models (LLMs), typically with over a billion parameters, have greatly advanced natural language processing across a wide range of domains. In the financial sector, LLMs offer promising capabilities for understanding complex documents, generating investment insights, and enabling data-driven decision-making. A growing number of specialized financial LLMs have been developed, including BloombergGPT~\cite{wu2023bloomberggpt}, DISC-FinLLM~\cite{chen2023disc}, and XuanYuan~\cite{zhang2023xuanyuan}, all of which are trained or adapted on corpora reflecting financial discourse and context.

Beyond general-purpose models, several financial LLMs have been tailored to support specific tasks and modalities. PIXIU~\cite{xie2023pixiu}, fine-tuned from LLaMA~\cite{2023arXiv230213971T}, targets structured financial tasks such as risk assessment and entity linking. FinVis-GPT~\cite{wang2023finvisgptmultimodallargelanguage} incorporates multimodal chart interpretation for visual-grounded analysis. InvestLM~\cite{yang2023investlmlargelanguagemodel} focuses on deep financial reasoning using curated QA datasets.

However, these models operate under a single-agent architecture, which limits their capacity to decompose and solve complex, multi-step financial tasks. This motivates exploration into multi-agent systems with modular and collaborative capabilities.

\subsection{Multi-Agent Collaboration}
\label{sec:2.2}

Multi-agent systems have emerged as promising solutions for tackling complex tasks. These systems employ strategies such as role-playing, collaboration, and task decomposition to improve problem-solving efficiency. For instance, AutoGen~\cite{wu2023autogenenablingnextgenllm} provides an open framework that enables agent communication for LLM-based applications, while MetaGPT~\cite{hong2023metagptmetaprogrammingmultiagent} adopts an assembly-line paradigm where specialized agents execute structured subtasks.

In the financial domain, recent studies have explored agent-based systems to support investment and trading decisions. TradingGPT~\cite{2023arXiv230903736L} models agents with different risk profiles and strategies. FinMem~\cite{yu2023finmemperformanceenhancedllmtrading} incorporates profiling, memory, and decision-making modules to improve cumulative returns. FinAgent~\cite{zhang2024multimodalfoundationagentfinancial} integrates image-based financial data into agent interactions to enhance trading decisions. However, most efforts remain trading-focused, lacking broader applications in macroeconomic, industry-level, and company-level financial analysis.

Recent work in other domains has demonstrated that multi-agent collaboration enhances performance in complex tasks. SMART~\cite{yue2025synergistic} leverages trajectory-based coordination to improve factual consistency in knowledge-intensive scenarios. MASER~\cite{yue2025multi} simulates legal interactions with role-aligned agents. In the medical domain, MDAgents~\cite{kim2024mdagents} and AI Hospital~\cite{fan2024ai} design adaptive agent collaborations for clinical reasoning and diagnosis, demonstrating gains in multi-turn interaction and decision-making. These works highlight the effectiveness of structured cooperation and role specialization, principles that motivate our multi-agent design for financial analysis.

\begin{figure}[t]
      \centering
      \includegraphics[width=0.8\columnwidth]{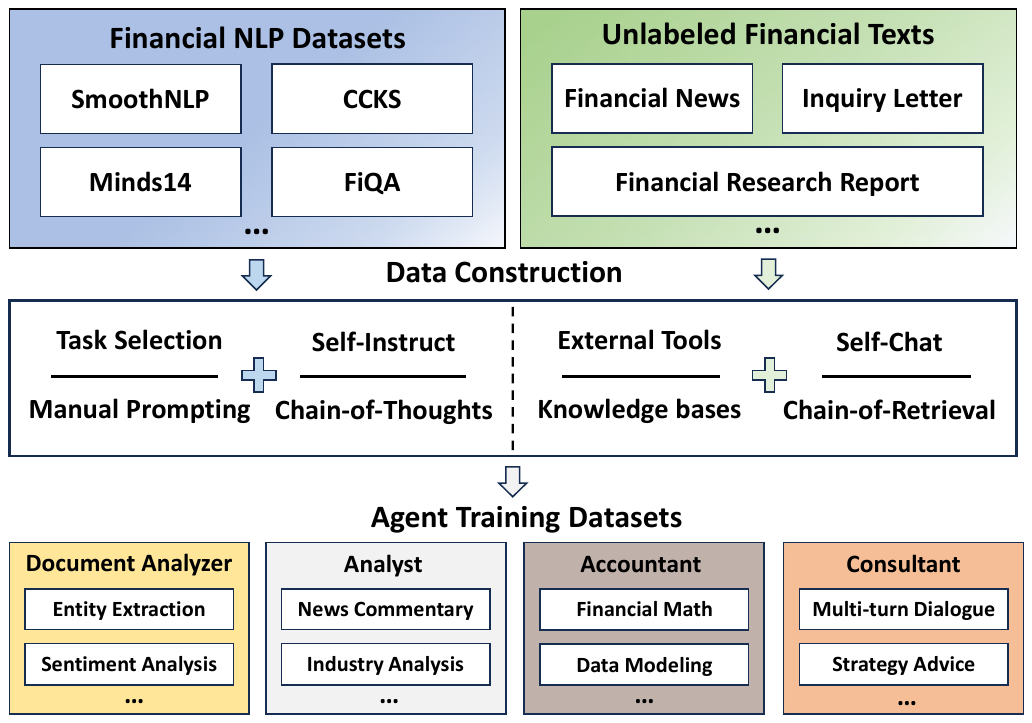}
      \caption{Construction of agent training dataset.}
      \label{fig:dataset}
\end{figure}

\section{FinTeam}
To meet the needs of practical financial scenarios, we propose a multi-agent collaborative financial intelligence system, FinTeam. This system organizes a virtual team of financial agents to handle complex tasks through agent interactions. Four roles are defined: \textit{document analyst}, \textit{analyst}, \textit{accountant}, and \textit{consultant}, each specializing in specific financial skills via supervised training. The construction and statistics of the agent training dataset are shown in Figure \ref{fig:dataset} and Table \ref{tab:statistics}.
Users can deploy these agents individually to handle specific financial tasks or enable collaboration across three main scenarios ~\cite{robbins2011management}—macroeconomic, industry, and company analysis—to tackle complex financial challenges. The overall overview of our system is shown in Figure \ref{fig:finllm}.

\subsection{Agent Roles}
\label{sec:3.1}
\subsubsection{Document Analyzer}
The financial industry generates vast amounts of data daily, particularly unstructured text data such as news, market commentaries. 
In our system, the \textit{document analyzer} serves as an agent designed to processing financial texts, capable of performing NLP tasks such as intent recognition, financial entity extraction, and financial sentiment analysis, tailored to specific financial texts and user requirements. 

We use domain-specific NLP datasets to train the \textit{document analyzer}. 
The dataset comprises two parts: labeled open-source datasets and unlabeled texts in financial reports that are automatically annotated by ChatGPT. The details of NLP
datasets is shown in Table \ref{tab:nlp_datasets} in Appendix \ref{sec:appendix1}.

\subsubsection{Analyst}

The financial field evolves rapidly, requiring analysis in context of current events. Our system's \textit{analyst} uses Retrieval-Augmented Generation (RAG) to analyze real-time financial materials—news, policies, reports, and data—retrieved from multiple knowledge bases via the m3e-base~\cite{Moka} embedding model.

We propose the Chain-of-Retrieval (CoR) prompting methodology to create a financial analysis instruction dataset, consisting of three main steps:
\textbf{1) Question Generation:} Formulate financial analysis questions based on financial contexts. 
\textbf{2) Reference Retrieval:} Retrieve relevant documents from knowledge bases. 
\textbf{3) Answer Generation:} Combine questions with retrieved contexts to generate answers. This dataset includes financial questions, reference documents, and their corresponding analyses, all generated by ChatGPT.

\subsubsection{Accountant}

Financial texts, particularly reports, often contain complex numerical data, requiring calculations like growth rates or expected earnings. Since LLMs struggle with accurate calculations~\cite{golkar2023xvalcontinuousnumberencoding}, we introduce accountant, a tool-augmented agent that utilizes computational tools to address this limitation, following the Toolformer approach~\cite{Toolformer}. 

\textit{Accountant} is equipped with four tools, as shown in Table~\ref{2}, which cover various computing tasks with different commands, inputs, and outputs. For instance, the calculator command is [Calculator(expression)→result]. When a quantitative analysis query is received, accountant generates the appropriate tool call command, halts decoding, executes the calculation, and integrates the result into the response before continuing with the text generation.

For the training set construction of \textit{accountant}, we create a seed task pool with three components: financial calculation questions from exams, arithmetic questions within financial report contexts, and general math questions from BELLE School Math~\cite{ji2023exploring}. The answers to these questions include embedded tool commands specifying their usage. Additionally, we use ChatGPT~\cite{openai2023chatgpt} to generate over 50,000 new question-answer pairs using self-instruction~\cite{selfinstruct} and few-shot Chain-of-Thought (CoT) prompting, with answers incorporating calculation tool commands.

\renewcommand{\dblfloatpagefraction}{.8}
\vspace{-1em}
\begin{table}[H]
    \centering
    \small
    \resizebox{\columnwidth}{!}{
    \begin{tabular}{ll} \toprule
    
        \scshape Tool             &  \scshape Detail  \\ \midrule
    Expression calculator     & Input: expression; Output: result\\\midrule
    Equation solver &Input: equation system; Output: solution \\ \midrule
    Counter       & Input: array of samples; Output: sample size              \\ \midrule
    Probability table  & Input: number; Output: cumulative standard normal distribution function value at this number\\
    \bottomrule
    \end{tabular}
    }
\caption{Definition of calculation tools}
\label{2}
\end{table}

\subsubsection{Consultant}

The finance domain is vast, covering areas like banking, stock trading, and futures. A financial intelligence system needs extensive background knowledge to perform detailed analysis. To address this, we developed an agent called \textit{consultant}, designed to interpret and answer finance-related queries.

To equip the \textit{consultant} with solid financial knowledge and consulting skills, we construct a Chinese-language dataset. We begin by translating the FiQA dataset~\cite{maia201818} and use only the Chinese version for training. To enhance domain understanding, we generate QA pairs from 200 finance-specific terms using ChatGPT~\cite{openai2023chatgpt}. We also scrape user queries from the Chinese financial forum JingGuan\footnote{\url{https://bbs.pinggu.org/}}, and apply a self-chat approach~\cite{xu2023baize} to create multi-turn dialogues based on seed topics.

\begin{table}[H]
\centering
\small
\resizebox{0.63\columnwidth}{!}{
\begin{tabular}{rrrr}
\toprule
\scshape Agent & \scshape \#Samples & \scshape Input Length & \scshape Output Length \\ \midrule

\textit{Consultant} & 63k        & 26   & 370 \\
\textit{Document Analyzer}  & 95k        & 586  & 31  \\
\textit{Accountant} & 64k        & 70   & 204 \\
\textit{Analyst} & 20k        & 1023 & 521 \\ 
\midrule
Total & 241k        & 340  & 206 \\
\bottomrule
\end{tabular}}
\caption{Statistics of agent training dataset. The lengths are the average number of tokens.}
\label{tab:statistics}
\end{table}

\subsection{Scenario Settings}
\label{sec:3.2}

To provide comprehensive financial analysis, we design three core scenarios—ranging from macro to micro levels—illustrated in Figure~\ref{fig:finllm}. These scenarios enable flexible, tailored responses to user needs.

\textbf{Macroeconomic Analysis.}
This scenario focuses on macroeconomic theories and trends. FinTeam operates as follows: 1) the \textit{document analyzer} extracts key terms; 2) the \textit{consultant} explains them; 3) the \textit{analyst} gathers and summarizes supplementary data; 4) the \textit{consultant} compiles a final response. This process helps users stay informed about economic developments and make better investment decisions.

\textbf{Industry Analysis.}
This scenario addresses specific sectors and trends. FinTeam proceeds as follows: 1) the \textit{document analyzer} identifies relevant industries or companies; 2) the \textit{analyst} explores competition, supply chains, and development trends; 3) recent news is included if requested; 4) the \textit{consultant} summarizes insights and offers strategic suggestions. Users gain a clear understanding of industry dynamics and outlook.

\textbf{Company Analysis.}
This scenario evaluates individual companies using models like PEST and SWOT. FinTeam works as follows: 1) the \textit{document analyzer} extracts key company data; 2) the \textit{analyst} applies PEST, SWOT, and performs sentiment analysis if needed; 3) the \textit{consultant} delivers a synthesized assessment.

Additionally, this scenario includes \textbf{financial statement analysis}: 1) the \textit{analyst} retrieves data from balance sheets, income statements, and cash flow reports; 2) the \textit{accountant} calculates key ratios such as profitability, liquidity, and leverage; 3) the \textit{consultant} generates a concise, actionable report.

Overall, this scenario provides structured, in-depth insights into a company’s financial health, market position, and strategic outlook, supporting informed investment decisions.

\section{Experiments}
\subsection{Evaluation Data and Setups}

To assess the performance of our FinTeam in real-world scenarios, we gather 150 actual investor inquiries from the popular Chinese online investment forum, NGA Grand Era~\footnote{\url{https://ngabbs.com/}}. For the three major scenarios, we ensure diversity in the collected questions by applying rule-based filtering, ultimately retaining fifty test questions per scenario.

In the macroeconomic scenario, questions address popular topics such as changes in economic indicators, asset price fluctuations, market interest rate variations, and global financial policy news. The industry scenario encompasses inquiries from 27 sub-sectors, including industry news evaluations and investment trends. The company scenario focuses on highly followed publicly listed companies, involving news, earnings reports, and stock price fluctuations.

For evaluation, we employ GPT-4o to score the outputs from our agent system and the other models, ensuring objectivity and accuracy. The evaluation is conducted across four dimensions:
\textbf{(1) Accuracy}: The model addresses key points directly, avoiding irrelevant details.
\textbf{(2) Thoroughness}: The model provides a detailed, in-depth answer.
\textbf{(3) Clarity}: The response is clear, concise, and logical.
\textbf{(4) Professionalism}: The model uses appropriate financial perfessional terms.

We compare model outputs across multiple dimensions, with GPT-4o rating each response from 1 to 5 per category and assigning an overall score. Pairwise significance tests are conducted to confirm statistical improvements. To validate GPT-4o’s judgments, human evaluations are also performed by finance undergraduates. They anonymously assess responses and select the best ones. The Acceptance Rate indicates how often a model’s output is chosen as the top answer.

\subsection{Implementation Details}

We train the agents using the LoRA mechanism~\cite{hu2021loralowrankadaptationlarge} on the Qwen2.5-7B-Chat model with Deepspeed ZeRO-0~\cite{2023arXiv230610209W} on four NVIDIA V100-32G GPUs. The batch size is 1 per GPU, with gradient accumulation steps of 4, a maximum sequence length of 4096, and 2 epochs. The learning rate is \(5 \times 10^{-5}\) and follows a cosine annealing schedule. LoRA parameters are set with a target of "all", rank 8, and alpha 16.

\subsection{Main Results}

We compare performance of FinTeam with the baseline models, with results presented in Table \ref{tab:comparison}. From these evaluations, we observe that our model achieve a 0.13 improvement in overall score when answering Chinese financial questions, compared to the baseline. The most significant gains are in thoroughness and professionalism, where the model improve by 0.23 points in both categories. Additionally, our financial agent collaboration system outperform GPT-3.5-turbo and Xuanyuan-13B across all dimensions, demonstrating its effectiveness.The results of significance tests, presented in Table \ref{tab:statistics_43}, confirm that the improvements in thoroughness, professionalism, and overall score are highly significant, with p-values well below the accepted thresholds, demonstrating the robustness of our system’s performance.
\begin{table}[H]
\centering
\small
\resizebox{0.55\columnwidth}{!}{
\begin{tabular}{lccccc}
\toprule
\scshape Model    & \scshape Acc. & \scshape Tho. & \scshape Cla. & \scshape Pro. & \scshape Overall \\ \midrule
\textbf{FinTeam (ours)}    & 4.54  & \textbf{4.94}  & 4.84  & \textbf{4.96}  & \textbf{4.86}  \\
\textbf{Qwen2.5-7B-Chat}     & 4.51  & 4.69  & \textbf{4.99}  & 4.80 & 4.78  \\
\textbf{GPT-4o}      & \textbf{4.67}  & 4.73  & 4.99  & 4.85  & 4.83  \\
\textbf{ChatGLM3-6B}      & 3.95  & 3.75  & 4.68  & 3.91  & 4.04  \\
\textbf{Xuanyuan-13B} & 4.35  & 4.61  & 4.96  & 4.67  & 4.66  \\ \bottomrule
\end{tabular}
}
\caption{Evaluation results across different dimensions. FinTeam achieves the highest overall score, surpassing the baseline model by 0.08, with notable improvements of 0.25 in thoroughness and 0.16 in professionalism. These results highlight the effectiveness of FinTeam.}
\label{tab:comparison}
\end{table}
\vspace{-2.5em}
\vspace{-1em}
\begin{table}[H]
    \centering
    \small
    \resizebox{0.5\columnwidth}{!}{
    \begin{tabular}{lccccc}
    \toprule
    \scshape Metric    & \scshape Acc. & \scshape Tho. & \scshape Cla. & \scshape Pro. & \scshape Overall \\ \midrule
    \textbf{t-statistic} & 0.253  & 5.445  & -5.195  & 4.218  & 2.493  \\
    \textbf{p-value}     & 0.8005  & \textbf{0.0000}  & \textbf{0.0000}  & \textbf{0.0000}  & \textbf{0.0138}  \\ \bottomrule
    \end{tabular} }
    \caption{Statistical significance of model evaluation metrics. The results indicate that the improvements in thoroughness, professionalism, and overall score are statistically significant.}
    \label{tab:statistics_43}
\end{table}
\vspace{-2em}
The final count of human evaluation picks is shown in Table \ref{tab:comparison2}. 
FinTeam significantly outperforms the other models, achieving an acceptance rate of 62.00\%. The results are consistent with the GPT-4o assessment, further validating our system's reliability.
It can be concluded that FinTeam is capable of providing professional and thorough answers to users' questions in real financial scenarios, offering users an in-depth understanding across multiple materials.
\vspace{-1em}
\begin{table}[H]
\centering
\small
\resizebox{0.85\columnwidth}{!}{%
\begin{tabular}{lccccc}
\toprule
Metric & FinTeam (ours) & Qwen2.5-7B-Chat & GPT-4o & ChatGLM3-6B & Xuanyuan-13B \\ \midrule
Acceptance Rate & \textbf{62.00\%} & 9.33\% & 5.33\% & 4.00\% & 19.33\%\\
\bottomrule
\end{tabular}}
\caption{Human evaluation on the preference of model outputs. Acceptance Rate indicates how often a model's answer is selected as the best among all models. FinTeam is chosen 93 times out of 150 test cases, achieving a selection rate of 62.00\%.}
\label{tab:comparison2}
\end{table}
\vspace{-2em}

\section{Analysis}
What causes our FinTeam to give better generation quality? 
We specifically analyze how each LLM agent performs in three different aspects. 

\renewcommand{\dblfloatpagefraction}{.8}

\subsection{Evaluation Setup}

\paragraph{Financial NLP Tasks}
We assess the model's NLP ability using the FinCUGE benchmark~\cite{2023arXiv230209432L} across six tasks: sentiment analysis (FinFE), event entity (FinQA), causality extraction (FinCQA), summarization (FinNA), relation extraction (FinRE), and entity extraction (FinESE). We create a few-shot evaluation setting with prompts and measure performance using accuracy, F1 score, and ROUGE score.

\paragraph{Chinese Financial Knowledge Tests} ~ {To evaluate our model's performance on Chinese financial knowledge, we utilize the FinEval~\cite{zhang2023fineval}, which covers 34 financial subcategories, containing a total of 1,151 multiple-choice questions. 
FinEval is Out-of-Distribution for our dataset, so it can adequately assess the generalization ability of our model and dataset. We measure the performance by calculating the accuracy of multiple-choice questions. 
}

\paragraph{Data Analysis}
We manually created a dataset of 100 financial calculation problems, adapted from material analysis questions in the Chinese Administrative Aptitude Test, to evaluate our model's capabilities. The dataset is crafted for quality assurance, and the model’s performance is assessed based on accuracy in formula construction and result calculation.

\subsection{Results on Financial NLP Tasks}

Table.\ref{tab:finbench_comparison_full} presents the performance of various models across six financial NLP tasks. The Document analyzer demonstrates the highest performance in all task-specific metrics. With an average score of 47.20, it significantly outperforms the strong baseline Qwen2.5-7B-Instruct, which achieves an average of 39.77. This represents a notable 7.43-point improvement, underscoring the effectiveness and robustness of the Document analyzer in handling diverse financial text understanding and reasoning tasks. 

\vspace{-1em}
\begin{table}[H]
\centering
\small
\resizebox{1\columnwidth}{!}{
\begin{tabular}{lccccccc}
\toprule
\scshape Model & \scshape FinFE (Acc) & \scshape FinQA (F1) & \scshape FinCQA (F1) & \scshape FinNA (ROUGE) & \scshape FinRE (Acc) & \scshape FinESE (F1) & \scshape Avg \\ \midrule
\textbf{Document analyzer} & \textbf{66.99} & \textbf{47.44} & \textbf{46.12} & \textbf{41.10} & \textbf{22.36} & \textbf{61.19} & \textbf{47.20} \\
\textbf{Qwen2.5-7B-Instruct}       & 66.19 & 37.33 & 34.50 & 40.80 & 21.02 & 36.76 & 39.77 \\
\textbf{ChatGLM3-6B}       & 62.57 & 25.69 & 21.41 & 30.60 & 8.39  & 28.44 & 29.18 \\
\textbf{Xuanyuan-13B}      & 62.48 & 18.07 & 24.94 & 31.10 & 9.13  & 31.08 & 29.47 \\ \bottomrule
\end{tabular}
}
\caption{Performance comparison across six financial tasks on FinCUGE benchmark.\textit{Document analyzer} outperforms the base model Qwen2.5-7B-Instruct by 7.43 points on average.}
\label{tab:finbench_comparison_full}
\end{table}
\vspace{-2.5em}

\subsection{Results on Financial Knowledge Tests} 

Table \ref{tab:fineval_avg} shows the evaluation results of our four LLM agents compared to general and financial LLMs on the FinEval benchmark. This demonstrates their extensive financial knowledge, strong task performance, and adaptability across diverse financial scenarios. Additionally, since FinEval is an out-of-distribution for our dataset, it highlights the universality of our training tasks and dataset.

\vspace{-1em}
\begin{table}[H]
\centering
\small
\resizebox{1\columnwidth}{!}{\begin{tabular}{lccccc}
\toprule
\scshape Model & \scshape \textit{Consultant} & \scshape Qwen2.5-7B-Instruct & \scshape ChatGLM3-6B & \scshape GPT-4o & \scshape Xuanyuan-13b \\ \midrule
\textbf{FinEval(Acc)} & 68.48 & 66.42 & 47.64 & \textbf{70.67} & 35.15 \\ \bottomrule
\end{tabular}
}
\caption{Experimental results of average scores on the FinEval benchmark. \textit{Consultant} outperforms the base model Qwen2.5-7B-Instruct by 2.06 points.}
\label{tab:fineval_avg}
\end{table}
\vspace{-1.5em}

\subsection{Results on Data Analysis} 
Table. \ref{tab:computing} showcases the experiment results on financial computing tasks. The addition of computational plugins to our model generates a notable performance boost compared to the baseline models, surpassing it by 0.09 in formula\&results. These results highlight the efficacy of our approach in addressing computational challenges within the financial domain. 
\begin{table}[H]
\centering
\small
\resizebox{0.6\columnwidth}{!}{
\begin{tabular}{lcc}
\toprule
               \scshape Model    & \scshape Formula &  \scshape Formula \& Result  \\ \midrule
\textit{Accountant}      & 0.70      & 0.70               \\ 
Qwen2.5-7B-Instruct & 0.65      & 0.61                \\ 
GPT-4o     & 0.81      & 0.81               \\

\bottomrule
\end{tabular}}
\caption{Evaluation results of financial calculation. Formula represents the accuracy of the formula in the calculation process, while Formula\&Result denotes the accuracy of both formulas and results. \textit{\textit{Accountant}} is 0.09 higher than the base model in Formula\&Result accuracy.
}
\label{tab:computing}
\end{table}

\section{Conclusion}
In this paper, in order to meet the needs of users in actual financial scenarios, we propose a financial intelligence system FinTeam, which connects multiple subtasks through the interaction between LLM agents for enhancing the system's ability to handle complex tasks. We construct the agent training dataset and train four LLM agents using different sub-datasets, which enables the agents to complete complex financial tasks in three scenarios: macroeconomic analysis, industry analysis, and company analysis following collaborative workflows. Within the framework of financial evaluation, we establish multi-dimensional benchmarks and demonstrate their robust capabilities, underscoring the capability of FinTeam to offer substantial support across various financial scenarios.

\begin{credits}
\subsubsection{Limitations}
Our work has several limitations. The scenario design is limited in scope, leaving out many financial tasks for future exploration. As the system may generate investment-related advice, users should be cautious, as financial outcomes are not guaranteed. The system is also tailored to Chinese financial contexts, and its effectiveness in global markets remains untested.

\end{credits}

\section*{Acknowledgements}
This research is supported by the National Natural Science Foundation of China (Grant No.~62176058).  
The project’s computational resources are supported by the CFFF platform of Fudan University.
%
%
%
%
\bibliographystyle{splncs04}
\bibliography{custom}




\section{Appendix}
\label{sec:appendix}
\subsection{Information of NLP datasets}
\label{sec:appendix1}
To train the \textit{document analyzer} in NLP tasks, we use various datasets, including sentiment analysis, entity extraction, and summarization, to enhance its capabilities. The open-source datasets used are listed in Figure \ref{tab:nlp_datasets}. To ensure balance, we randomly sample 10,000 samples from subsets with over 10,000 samples.

\vspace{-1.5em}
\begin{table}[H]
\centering
\resizebox{0.5\columnwidth}{!}{
\begin{tabular}{llr} \toprule
Dataset            & Task Type              & \# Samples  \\ \midrule
FPB   ~\cite{malo2014good}             & Sentiment Analysis      & 18690                           \\
CCKS-NEC-2022  ~\cite{ccksnec2022}      & Causality Extraction    & 7499                            \\
SmoothNLP IEE \footnotemark[3]      & Event Extraction        & 3256                            \\
SmoothNLP NHG \footnotemark[3]    & Text Generation         & 4642                            \\
CCKS2022-event~\cite{ccks2022event}     & Event Classification   & 3578                            \\
Minds14~\cite{gerz2021multilingual}            & Intent Prediction   & 59143                           \\
Financial Report   & Entity Extraction       & 61705                           \\
OpenKG ~\cite{ren2022iree}            & Entity Extraction       & 7672                            \\
OpenKG             & Entity Extraction       & 67921                           \\
Wealth-alpaca-lora \footnotemark[4] & Keyword Generation      & 41825                           \\
\bottomrule                         
\end{tabular}
}
\caption{Data statistics of our financial NLP datasets.}
\label{tab:nlp_datasets}
\end{table}

\footnotetext[1]{\url{https://github.com/wwwxmu/Dataset-of-financial-news-sentiment-classification}}

\footnotetext[2]{\url{https://github.com/smoothnlp/FinancialDatasets}} 

\footnotetext[3]{\url{https://huggingface.co/datasets/gbharti/finance-alpaca}}

\end{document}